\documentclass[12pt] {report}

\usepackage{amsfonts}
\usepackage{amssymb}
\usepackage{amsmath}
\usepackage{fancyvrb}
\usepackage{epsfig}
\usepackage{graphicx} 
\usepackage{latexsym}
\usepackage{lscape}

\newtheorem{theorem}{Theorem} 
\newtheorem{lemma}[theorem]{Lemma}

\newtheorem{corollary}[theorem]{Corollary}
\newenvironment{proof}[1][Proof]{\begin{trivlist}  \item[\hskip \labelsep
{\bfseries #1}]}{\end{trivlist}}
\newenvironment{definition}[1][Definition]{\begin{trivlist} \item[\hskip
\labelsep {\bfseries #1}]}{\end{trivlist}}

\newcommand{\qed}{\nobreak \ifvmode \relax
\else  \ifdim\lastskip<1.5em \hskip-\lastskip \hskip1.5em plus0em
minus0.5em
\fi \nobreak \vrule height0.75em width0.5em depth0.25em\fi}

\def\lh{\hbox to 15pt{\vbox{\vskip 6pt\hrule width 6.5pt height 1pt} \kern
-4.0pt\vrule height 8pt width 1pt\hfil}}

\begin{document}

\begin{titlepage}

\vspace*{-2cm}

\vspace{.5cm}

\begin{centering}

\huge{The Maurer-Cartan structure of BRST differential }

\vspace{.5cm}

\large  {Jining Gao }\\

\vspace{.5cm}

 Department of Mathematics, North Carolina State University,
Raleigh, NC 27695-8205.

\vspace{.5cm}

\begin{abstract}

 In this paper, we construct a new sequence of generators of the BRST
complex and reformulate the BRST differential so that it acts on elements of the complex much like the Maurer-Cartan differential acts on left-invariant forms. Thus our BRST differential is formally analogous to the differential defined on  the BRST formulation of the Chevalley-Eilenberg cochain complex of a Lie algebra. Moreover, for an important class of physical theories, we show that in fact the differential is a Chevalley-Eilenberg differential. As one of the applications of our formalism, we show that the BRST differential provides a mechanism   which permits us to extend a nonintegrable system of vector fields on a manifold to an integrable system on an extended manifold.
 \end{abstract}

\end{centering}

\end{titlepage}

\newpage

\section{Introduction}

Homological algebra has become an indispensable tool for the rigorous
formulation of a wide variety  of developments in theoretical physics. Applications of
these techniques to physics has become so pervasive that  they have gradually become identified as a new category of mathematical physics which has been called ``cohomological physics". One of the fruitful branches of this theory is the ``cohomology" formulation of the BRST theory of constraints. Indeed the point of BRST theory is to replace the cohomology of the reduced space of a physical theory by the cohomology of a homological resolution of the space $P$ being constrained.

In more detail, assume that $P$ is a symplectic  manifold and that one has a system of first class constraints on $P.$ Let $\Sigma$ denote the constraint surface defined as the set of zeros of the constraints. These constraints may or may not be independent. If they are independent they are called irreducible constraints and otherwise they are reducible. The Hamiltonian vector fields of the symplectic manifold $P$ define a possibly singular foliation of $\Sigma$ and the smooth functions on $\Sigma$ which are constant on the leaves of this ``foliation" are said to be gauge invariant and are called the observables of the theory. There is a differential $d,$ called the longitudinal differential, defined on a certain (dual) Chevalley-Eilenberg  complex with coefficients in the algebra $C^{\infty}(\Sigma)$ whose cohomology in degree zero in the irreducible case is the space of observables.  It is clear in the literature that the  zero degree cohomolgy of a certain complex is the !
 space of observables but it is not clear that the complex is a Chevalley-Eilenberg complex and that the longitudinal differential is a Maurer-Cartan differential. These facts are established here in a rigorous manner for the first time.

BRST symmetry was developed in order to replace the original gauge
symmetry  on the constraint surface by a symmetry on the entire phase space $P$ in such a manner that the longitudinal differential $d$ could be extended to a new differential $S$ called the BRST differential to be defined on an enlarged complex in such a way that the BRST cohomology in degree zero is precisely the set of observables on $\Sigma.$ The procedure is nontrivial even in the irreducible case but even more convoluted in the reducible case. An interesting question has to do with whether the BRST differential is a Maurer-Cartan differential and whether or not it is actually a Chevalley-Eilenberg differential defined on a (dual) Chevalley-Eilenberg complex as was the case for the longitudinal differential in the irreducible case.

The first chapter is mainly devoted to showing that in the case of irreducible constraints the BRST differential $S$ is in fact a  Maurer-Cartan differential and  that it is a (dual) Chevalley-Eilenberg differential defined on a Chevalley-Eilenberg complex.  It is also shown that in the case one has a Hamiltonian system subject to Bosonic irreducible constraints that the fact that $S^2=0$ implies the existence of a possibly singular ``foliation" of the phase space $P$ which agrees with the ``foliation" of the constrained space defined by gauge symmetries. Generally, the BRST differential has an expansion
$$ S=\delta +d+s_1 +\cdots +s_n+\cdots.$$
The Koszul Tate differential   $\delta$ and the longitudinal differential $d$ are well understood but the other terms of the expansion are less well understood. We completely characterize $s_1$ in the irreducible case. Finally, in this chapter we also consider systems whose constraints are reducible. In particular we introduce a new concept which we call an $n$th-reducible complex. This is precisely the idea needed to formulate reducible physical theories rigorously. 
We show that every differential on such a complex is a mildly generalized Maurer-Cartan differential. In particular we show that the BRST differential is such a generalized Maurer-Cartan differential in the reducible case.

\section{The Maurer-Cartan structure of BRST differential under the irreducible constraints}
Let $(P,\omega)$ be a $n$- dimensional symplectic manifold  and let $[,]$ be
the Poisson bracket defined by $\omega$ on the algebra of smooth functions 
$C^{\infty}(P).$ Assume that $G_a ,
a=1,\cdots,M$ are   constraint functions which satisfy the condition :
 
\begin{eqnarray}
[G_a,G_b]=C_{ab}^{c} G_c  \label{l2}
\end{eqnarray}
where $C_{ab}^{c}$ are structure functions on $P$ and
let $\Sigma$ be the constraint surface which is determined by the set of zeros of 
$G_a.$ When (\ref{l2}) is satisfied  we say that the constraints $G_a$ are first class
constraints. The Hamiltonian vector fields $X_a$ corresponding to the functions $G_a$ are defined 
by $X_a (f)=[f,G_a]$ for $f \in C^{\infty}(P).$ The fields $X_a$ satisfy
the condition $[X_a,X_b]\approx C_{ab}^c X_c,$  i.e.,  the equation holds only
on $\Sigma ,$  or as we say, they hold only "on shell". Under certain 
conditions the fields  $X_a$ define a foliation of $\Sigma.$ Functions 
$ f\in C^{\infty}(\Sigma)$ which are constant on the leaves of the 
foliation are said to be "gauge invariant" and are called classical 
observables. 
 
 In quantum
field  theory, it is difficult to utilize  path integrals of functionals defined on the 
space of  observables because they are only defined on the constraint surface.
To overcome this difficulty the phase space $P$ is extended and the gauge 
symmetry is replaced by BRST symmetries  in such way that the path integral can be utilized   on functionals defined on arbitrary functions on the
extended phase space. More precisely, to achieve this, 
one introduces an antighost variable $P_a$ for every constraint function $G_a$
and a differential $\delta$ called  the Koszul-Tate 
differential which is defined on the complex $C[P_a]\otimes C^{\infty}(P)$ as follows:
\begin{eqnarray}
\delta P_a =-G_a \\ \delta f=0 \\ \delta \eta^a=0
\end{eqnarray}
where $f\in C^{\infty}(P)$. Additionally, new variables $\eta^a$ are introduced which are in one-to-one correspondence with the space of independent 
gauge symmetries and another differential $d$ called the longitudinal differential is defined  on the complex $C^{\infty}(P)\otimes C[\eta^b]$ in a manner similar to the definition of the Chevalley-Eilenberg differential. This differential is designed to implement the gauge symmetries. In some cases $\delta +d$ is a differential on the complex $C[P_a]\otimes C^{\infty}(P)\otimes C[\eta^b]$ whose square is zero and whose cohomology is precisely the space
of classical observables. Often this fails to be true and $\delta+d$ must  be 
extended by homological perturbation theory to obtain the so-called BRST 
differential 
$S=\delta + d +S_1+\cdots $
in order to obtain the observables as  zero degree cohomology classes. The differential $S$ is clearly quite
different from the longitudinal differential $d,$ but we will
show that  $S$ satisfies conditions totally analogous to those characterizing $d$  in Henneaux and Teitelboim  (\cite{HT} page117-119) in the case when the constraint functions are irreducible and Bosonic.

Let $\Omega=C[P_a]\bigotimes C^{\infty}(P)\bigotimes C[\eta^b]$ and consider 
$ \Omega^{*}=\bigoplus_{p=0}^{\infty}\Omega^p $, where $\Omega^{p}$ is the subset of $\Omega$ having
ghost number $p$ (defined below).  For simplicity,
we introduce the notation 
$\omega^I =\eta^{b_1}
\cdots \eta^{b_{p+1}}P_{a_p} \cdots P_{a_1} $, and 
$\Omega^1= \{\alpha \in \Omega \mid \alpha =u_I \omega^I ,u_I \in C^{\infty}(P)\}$
where $I$ is the multi-index $(b_1,\cdots,b_{p+1}, a_1,\cdots,a_{p}).$
Obviously the elements $ \omega^I$
generate all $\Omega^p$ for $p\geq 1.$ For completeness and clarity,
we first describe our parity conventions  as follows:
\begin{eqnarray}
\epsilon(P_a)=\epsilon(G_a)+1=\epsilon(\eta^a), \epsilon(AB)=\epsilon(A)+\epsilon(B).
\end{eqnarray}
Moreover  the ghost number grading referred to above is defined as follows:
\newline
\indent (1) the pure ghost number of each element of $\Omega$  is simply its degree as a polynomial  in $\eta^a,$

(2) the anti- ghost number of each element of $\Omega$ is its degree as a polynomial  in $P_a,$

(3) the ghost number of each element $x$ is the number  $puregh(x)- antigh(x).$
\newline
Notice that for $A,B\in \Omega,$ gh($AB$)=gh($A$)+gh($B$) 
and that  $\epsilon (\omega^I)=1$ whenever $\epsilon (G_a)=0.$ With these conventions, we will show that the BRST
differential is  essentially the Chevalley-Eilenberg differential
when the constraints are Bosonic and irreducible.

First, we  recall how the Chevalley-Eilenberg differential is formulated in BRST notation.
Let ${\cal G}$ be a Lie algebra spanned by a basis $\{ e_i \}$ and ${\cal A}$ a  commutative associative algebra .
Let the mapping $\rho: {\cal G}\rightarrow End({\cal A})$ be  a representation of ${\cal G}$ with representation 
space ${\cal A}.$  Introduce a ghost variable $\eta^i$ for every element $e_i$ of the basis $\{ e_i \}$.
Let $ A$ denote the $Z$-graded algebra ${\cal A}\otimes C[\eta^1,\eta^2,\cdots ]$ with the grading defined by the ghost number.The Chevalley-Eilenberg differential $d=d_{CE}$ is defined on generators of the complex $A$  as follows: 
 \begin{eqnarray}
df=\rho(e_a)(f)\eta^a     \label {af}
\end{eqnarray}

\begin{eqnarray}
d\eta^a={-\frac{1}{2}}C_{cb}^{a}\eta^{b}\eta^{c} \label {a}
\end{eqnarray}
where $f,C_{ab}^{c} \in {\cal A}$ .  The mapping $d=d_{CE}$ is extended to the entire graded algebra $ A$ by the Leibniz law 
\begin{eqnarray}
d(\alpha \beta)=(d\alpha )\beta+(-1)^{deg \alpha}\alpha (d\beta) \label {la}
\end{eqnarray}
Any differential which satisfies the conditions (\ref {af}) and  ( \ref{a}) will be called a {\bf Maurer-Cartan differential}. Moreover we will say that $d$ is a  {\bf Chevalley-Eilenberg differential } whenever there exists a Lie algebra ${\cal G}$ and a representation $\rho$ into the endomorphisms of some commutative associative algebra ${\cal A}$ satisfying not only (\ref a) but also (\ref {af}) and (\ref {la}). We do not require that our  Lie algebra ${\cal G}$ be finite dimensional but our Lie algebras are finitely generated as modules over our algebra ${\cal A}.$

If we choose ${\cal A}=C^{\infty}(\Sigma)$ where $\Sigma$ is the constraint surface defined above and if
$e_a = X_a,$ then the longitudinal differential is  a Chevalley-Eilenberg differential of this type. In this case the vector fields $X_a$ must be restricted
to $\Sigma$ and the Lie algebra is the sub-algebra of vector fields on $\Sigma$ spanned by the $X_a$  over the algebra $C^{\infty}(\Sigma).$ The fact that this is a sub-Lie algebra follows from the identity $[X_a,X_b]=C_{ab}^dX_d+X_{C_{ab}^d}G_d.$ The other properties follow immediately. We want to obtain an ``off shell" version of this result.

Since $\epsilon (\omega^I)=1$ and $gh(\omega^I)=1$
we call  the set of monomials $\omega^I $ multi-ghosts. Moreover it follows from  $S\omega^I \in \Omega^2,$ that
\begin{eqnarray}
S\omega^K={-\frac{1}{2}}C_{IJ}^{K}\omega^I \omega^J
\end{eqnarray}
where $C_{IJ}^{K}=C_{JI}^{K}.$
Similarly, for $f \in \Omega^0,$   one has that $Sf=(\rho_I f)\omega^I,$ since $Sf \in \Omega^1.$
To summarize, we have following theorem:

\begin{theorem}
If the constraint functions $ \{G_a \} $ are irreducible and Bosonic, the relevant BRST differential $S$ defined on the complex $C[P_a]\otimes C^{\infty}(P)\otimes C[\eta^b]$ above is a  Maurer-Cartan differential:
\begin{eqnarray}
Sf=(\rho_I f)\omega^I  \label {rho}
\end{eqnarray}

\begin{eqnarray}
S\omega^K={-\frac{1}{2}}C_{IJ}^{K}\omega^I \omega^J.
\end{eqnarray}
\end{theorem}

Notice that  even though  the longitudinal differential $d$  is not nilpotent on the space
$C[P_a]\bigotimes C^{\infty}(P)\bigotimes C[\eta^b]$, its BRST extension  S is nilpotent and so is a
differential. To determine how the  BRST differential $S$ and the longitudinal differential $d$ are related,
we compare the following formulas with the formulas (\ref{af}) and (\ref{a})
$$
Sf=(\rho_I f)\omega^I=(\partial_a f)\eta^a+s_1 f+\cdots + s_n f+\cdots $$
\begin{eqnarray}S\eta^a ={-\frac{1}{2}}C_{cb}^{a}\eta^{b}\eta^{c}+s_1\eta^a+s_1\eta^a.  \label{b}
\end{eqnarray}
Note that the terms on the right hand sides of (\ref{af}) and  (\ref{a}) are summands of the right hand side of these equations.

We claim that the BRST differential $S$ is essentially a Chevalley-Eilenberg differential in the case that the constraints are Bosonic and irreducible. The required Lie algebra is a sub-Lie algebra of the Lie algebra $ {\cal X}(P)$ of all vector fields on $P.$ Since $S^2=0,$ each of the mappings $\rho_I$ defined by the equation (\ref {rho}) above is a derivation of $C^{\infty}(P)$ and so is a vector field on $P.$ We consider the submodule ${\cal G}(\rho)$ of  ${\cal X}(P)$ spanned by the vector fields $\rho_I$ over $C^{\infty}(P).$ We eventually show that it is a sub-Lie algebra of ${\cal X}(P).$ Each element of ${\cal G}({\rho})$  clearly acts as a derivation of the algebra ${\cal A}=C^{\infty}(P)$ and therefore is in $End({\cal A}).$ Once we establish the fact that ${\cal G}({\rho})$ is a Lie algebra we will have the required data in order to show that $S$ is a Chevalley-Eilenberg differential. First we need a lemma which is of interest in its own right.

\begin{lemma} Assume that the constraints are Bosonic and irreducible and consider the BRST differential $S$ on the complex  $C[P_a]\bigotimes C^{\infty}(P)\bigotimes C[\eta^b].$  Let $\rho=\rho_I$ denote the ``representation" defined by the identities in Theorem 1. Then
$$S^2 f=\frac{1}{2}([\rho_J,\rho_I]f-C_{JI}^K\rho_K f)\omega^J \omega^I.$$
Moreover if $S^2f=0$ for all $f\in C^{\infty}(P),$ then 
$$S^2\omega^K= -\frac{1}{6}\{[\rho_I,[\rho_J,\rho_E]]^K+[\rho_J,[\rho_E,\rho_I]]^K +[\rho_E,[\rho_I,\rho_J]]^K\}$$
\end{lemma}

\begin{proof}  First we prove the first identity. Notice first that since $S$ is an odd derivation, we have 
\begin{eqnarray}
S^2 f=S((\rho_I f)\omega^I)=S(\rho_I f)\omega^I +(\rho_I f)S\omega^I.
\end{eqnarray}
It follows from the identities of Theorem 1 that
\begin{eqnarray}
S^2 f=(\rho_J \rho_I) (f)\omega^J \omega^I+(\rho_I f)(-\frac{1}{2} C_{JK}^I\omega^J \omega^K) \\
=\frac{1}{2}[(\rho_J \rho_I) f-(\rho_I \rho_J) f]\omega^J \omega^I-\frac{1}{2}C_{JK}^I\omega^J \omega^K\rho_I f\\
= \frac{1}{2}[\rho_J,\rho_I]f\omega^J \omega^I-\frac{1}{2}C_{JI}^K\omega^J \omega^I\rho_K f \\
=\frac{1}{2}([\rho_J,\rho_I]f-C_{JI}^K\rho_K f)\omega^J \omega^I
\end{eqnarray}
Thus the first of the two identities is true.
We now prove the second identity. Since $ S\omega^{K}=-\frac{1}{2}C_{IJ}^{K}\omega^I\omega^J $, where $I=(b_1,\cdots,b_{p+1},a_1,\cdots,a_{p})$ and
$J=(\tilde{b}_1,\cdots,\tilde{b}_{p+1},\tilde{a}_1,\cdots,\tilde{a}_{p}),$
 we have
\begin{eqnarray}
S^2\omega^K=-\frac{1}{2}(SC_{IJ}^{K})\omega^I\omega^J -\frac{1}{2}C_{IJ}^{K}d\omega^I\omega^J
+\frac{1}{2}C_{IJ}^{K}\omega^I d\omega^J \nonumber \\
=-\frac{1}{2}(\rho_{E}C_{IJ}^{K})\omega^E\omega^I\omega^J-\frac{1}{2}C_{IJ}^{K}(-\frac{1}{2}C_{\tilde{K}L}^{I} \omega^{\tilde{K}} \omega^{L}) \omega^J   \nonumber \\
+\frac{1}{2}C_{IJ}^{K}\omega^I(-\frac{1}{2}C_{MN}^{J}\omega^M\omega^N)\nonumber \\
=-\frac{1}{2}\rho_{E}C_{IJ}^{K}\omega^E\omega^I\omega^J+\frac{1}{4}C_{IJ}^{K}C_{\tilde{K}L}^{I} \omega^{\tilde{K}} \omega^L \omega^J
-\frac{1}{4}C_{IJ}^{K}C_{MN}^{J}\omega^I\omega^M\omega^N \nonumber \\
=-\frac{1}{2}\rho_{E}C_{IJ}^{K}\omega^E\omega^I\omega^J+\frac{1}{2}C_{IJ}^{K}C_{\tilde{K}L}^I \omega^{\tilde{K}} \omega^L \omega^J \nonumber \\
=-\frac{1}{6}(\rho_{E}C_{IJ}^{K}+\rho_{I}C_{JE}^{K}+\rho_{J}C_{EI}^{K})\omega^I\omega^J\omega^E \nonumber\\
+\frac{1}{6}(C_{MI}^{K}C_{JE}^{M}+C_{MJ}^{K}C_{EI}^{M}+C_{ME}^{K}C_{IJ}^{M})\omega^I\omega^J\omega^E   \nonumber \\ -\frac{1}{6}(\rho_{E}C_{IJ}^{K}+\rho_{I}C_{JE}^{K}+\rho_{J}C_{EI}^{K})\omega^I\omega^J\omega^E \nonumber\\
 -\frac{1}{6}(C_{IM}^{K}C_{JE}^{M}+C_{JM}^{K}C_{EI}^{M}+C_{EM}^{K}C_{IJ}^{M})\omega^I\omega^J\omega^E   \label{h1}
\end{eqnarray}

Next notice that if we assume that $S^2f=0$ for all $f\in C^{\infty}(P)$, then $[\rho_J,\rho_E]=C_{JE}^M \rho_M $, $[\rho_E,\rho_I]=C_{EI}^M \rho_M $ , $[\rho_I,\rho_J]=C_{IJ}^M \rho_M $, and
we have 
\begin{eqnarray}
[\rho_I,[\rho_J,\rho_E]]=[\rho_I,C_{JE}^M \rho_M ]=(\rho_I C_{JE}^M) \rho_M +C_{JE}^M [\rho_I,\rho_M]\\
=(\rho_I C_{JE}^M) \rho_M+C_{JE}^M C_{IM}^K \rho_K \\
=(\rho_I C_{JE}^K +C_{JE}^M C_{IM}^K) \rho_K .
\end{eqnarray}
 
It follows that 
\begin{eqnarray}
[\rho_J,[\rho_E,\rho_I]]=(\rho_J C_{EI}^M) \rho_M+C_{EI}^M C_{JM}^K \rho_K \\
=(\rho_J C_{EI}^K +C_{EI}^M C_{JM}^K) \rho_K
\end{eqnarray}
and 
\begin{eqnarray}
[\rho_E,[\rho_I,\rho_J]]=(\rho_E C_{IJ}^M) \rho_M+C_{IJ}^M C_{EM}^K \rho_K \\
=(\rho_E C_{IJ}^K +C_{IJ}^M C_{EM}^K) \rho_K .
\end{eqnarray}
It follows from this last calculation that  the negative of the sum of  the $K$-th components of the right hand sides of the last three equations is precisely six times the  right hand side of  the identity for $S^2\omega^K$ (see (\ref {h1})).  The lemma follows. 
\end{proof}

\bigskip
\begin{corollary} Assume that the constraints are Bosonic and irreducible and consider the BRST differential $S$ on the complex  $C[P_a]\bigotimes C^{\infty}(P)\bigotimes C[\eta^b].$ Since in fact $S^2=0$ we have that ${\cal G}({\rho})$ is a Lie sub-algebra of ${\cal X}(P)$ with generators the set  of vector fields  $\{\rho_I\}$ on $C^{\infty}(P).$
\end{corollary}

\bigskip
\begin{corollary} If the constraints of a Hamiltonian system are Bosonic and irreducible, then the
BRST differential is a Chevalley-Eilenberg differential on the complex ${\cal A}\otimes C[\omega^I]$ where the algebra ${\cal A}$ is the algebra of smooth functions on $P$ and where the free generators $\omega^I$ are called multi-ghosts instead of ghosts.
\end{corollary}

\begin{proof} The proof was outlined in the observations just prior to the lemma. The only gap in the argument was that we had not yet proved that ${\cal G}(\rho)$ is a Lie algebra which we now see is a corollary of the lemma.

\bigskip
There is one caveat regarding the last Corollary and that is that in our definition of a Chevalley-Eilenberg differential the complex is ${\cal A}\times C[\eta^b]$ where the $\eta^b$ are free generators which are called ghosts. In the present case the $\omega^I$ are still free but the algebra $C[\omega^I]$  is a subalgebra of the algebra 
$C[P_a]\times C[\eta^b].$ The differential $S$ still qualifies to be called a Chevalley-Eilenberg differential however as we could merely rename the free generators $\omega^I$ and call them ghosts. We do not do this however due to the confusion which would arise preferring instead to call them multi-ghosts.
\end{proof}

\bigskip
{\bf Remark.} The longitudinal differential $d$ was initially defined ``on shell", that is to say the underlying manifold was the constraint surface $\Sigma.$  Formulated on this surface the square of $d$ is zero and its cohomology in degree zero is the set of classical observables. In order to use the path integral formalism it is useful to extend the formalism ``off shell". When the longitudinal differential $d$ is extended ``off shell" it no longer squares to zero and in fact the BRST differential was constructed to repair this defect.
The fact that $d$ squares to zero ``on shell" is related to the fact that the Hamiltonian vector fields $X_a$ close under Lie brackets ``on shell". They do not close ``off shell". The fact that the BRST differential squares to zero ``off shell" suggests that on should be able to supplement the vector fields $X_a$ with other fields to obtain an integrable system which ``foliates" $P$ in such a manner that the possibly singular leaves provides the ``foliation" of $\Sigma$ provided by the
Hamiltonian vector fields. We now show that this is true.
\bigskip

Recall that the generators $\rho_I$ of the Lie algebra ${\cal G}({\rho})$ correspond to the multi-ghosts $\omega^I =\eta^{b_1} \cdots \eta^{b_{p+1}}P_{a_p} \cdots P_{a_1} $ where  $I$ is the multi-index \newline $(b_1,\cdots,b_{p+1}, a_1,\cdots,a_{p}).$ In the case $p=0$ it is understood that $\omega^I$ is simply $\eta^b$ for some index $b.$ Thus the equation $S(f)=(\rho_If)\omega^I$ of Theorem 1 has the terms $(\rho_a f)\eta^a $ as certain of its summands. Recall that these terms correspond to the longitudinal differential d in the expansion
\begin{eqnarray}
 S=\delta +d+s_1 +\cdots +s_n+\cdots. \label{d1}
\end{eqnarray}
of the BRST differential. Indeed 
for every  $f\in C^{\infty}(P),$ 
\begin{eqnarray}
  Sf=(\rho_a f)\eta^a +(\rho_I f)\omega^I+\cdots \label{c1}
\end{eqnarray}
where antidegree($\omega^I) \geq 1,$ and we see that  $\rho_a f$ is exactly the action of $X_a$ on $f.$ 
Consequently the $\rho_a$ are simply the Hamiltonian vector fields $X_a.$ The supplementary vector fields we require to obtain an integrable system are defined by $X_I=\rho_I.$ 
 
 As an immediate consequence of these observations we have the following theorem.
  
\begin{theorem}
 Let $d$ be the longitudinal exterior differential which, by construction of  the BRST operator $S$ is one of the summands in the expansion of $S$: $ S=\delta
+d+s_1 +\cdots +s_n+\cdots $ where $s_k$ is a derivation which increases the antighost degree by $k$. The longitudinal differential $d$ is defined in terms of the Hamiltonian vector fields  ${X_i}$ which form an open gauge algebra   since 
$[X_i,X_j]\approx C_{ij}^k X_k $. Since $S^2=0,$  there exists  extended vector fields ${X_I}$ on $P$ such that
$[X_i,X_j]= C_{ij}^k X_k+C_{ij}^I X_I $, and the fields ${X_i, X_I}$ define  an integrable system in the sense that they generate a subalgebra ${\cal G}(\rho)$ of the Lie algebra of all vector fields of $P.$
\end{theorem}

We now determine further conditions imposed on the $\rho_K$ by the fact that  $ S$ is nilpotent.
 
Using (\ref{d1}) and the fact that $S^2 =0,$ the first three terms of the expansion of 
$ S^2$ in terms of the anti-ghost degree yields : 
\begin{eqnarray}
&& \delta^2 =0 \\&& [\delta, d]=0 \\ && d^2 =-[\delta,s_1]
\end{eqnarray}
By a calculation similar to the one in the proof of the lemma, we have  
\begin{eqnarray}
d^2 f=\frac{1}{2}([\rho_i,\rho_j]f-C_{ij}^k\rho_k f)\eta^i \eta^j \label{e}
\end{eqnarray}
and for arbitrary  $f \in C^{\infty}(P)$
\begin{eqnarray}
[\delta, s_1]f=(\delta s_1 f+s_1\delta f) \\
=\delta s_1 f=\delta(\rho_{ab}^c f\eta^a\eta^b P_c)
\end{eqnarray}
where the $\rho_{ab}^c$ are defined by the equation $s_1f=\omega^If=\rho_{ab}^c f\eta^a\eta^b P_c$ and the multi-index $I$ is  $(abc).$
It follows that
\begin{eqnarray}
[\delta, s_1]f=(\rho_{ab}^c f)\eta^a\eta^b\delta P_c\\
=-(G_c)(\rho_{ab}^c f)\eta^a\eta^b \label{f}
\end{eqnarray}

Using the three identities above and comparing  (\ref{e}) with (\ref{f}), we have

\begin{eqnarray}
[X_j,X_i]= C_{ji}^k X_k+G_c\rho_{ij}^c \label{f1}
\end{eqnarray}
Since $\omega^I$ is a derivation for each multi-index $I$ we see that each $\rho_{ij}^c$ is  also a derivation on $C^{\infty}(P);$  we  distinguish
it from  the  derivation $\rho_k$ by referring to it as a second order derivation.
\bigskip

We now show how these results may be applied to Hamiltonian systems having
 first-class constraints restricting our remarks to the case where $P$ is $R^n$ for some positive integer $n.$  We adopt the same conventions as in \cite{HT} (Page 52-53), in particular, let 
$G_a (a=1,\cdots,M) $ denote the  constraint functions of the system. Define vectors  
$X_a=(X_a^{\lambda})$ via $X_a^{\lambda}=\sigma^{\lambda \mu}\partial_{\mu}G_a, $
and observe that 
 
\begin{eqnarray}
X_a^{\lambda}\partial_{\lambda}F=X_a F=[F,G_a].
\end{eqnarray}
Here the matrix of components of the antisymmetric tensor $(\sigma^{\lambda \mu} )$ is the inverse of the matrix $\omega_{\mu\nu}$ of components of the symplectic structure $\omega$ on $P.$
We know that if $X_a^{\lambda}$ corresponds to $G_a $ and $X_b^{\lambda}$ corresponds to $G_b $
then $ [X_a,X_b]^{\lambda} $ corresponds to $[G_a,G_b] $.
Moreover
\begin{eqnarray}
[X_a,X_b]^{\lambda}=\sigma^{\lambda \mu}\partial_{\mu}(C_{ab}^c G_c)\\
=C_{ab}^c X_c^\lambda+G_c \sigma^{\lambda \mu}\partial_{\mu}C_{ab}^c \approx C_{ab}^c X_c^\lambda \label{k1}
\end{eqnarray}

Off the constraint surface, the second term on the right hand side of (\ref{k1}) does not vanish unless
$\partial_{\mu}C_{ab}^c =0$. Thus $X_a^{\lambda}  (a=1,\cdots, M) $ form a closed distribution only on
shell $G_a=0.$

\bigskip

For the remainder of this section we provide a detailed calculation which shows how to determine the summand $s_1$ of the expansion of the BRST operator $S.$
\bigskip

First we determine the action of $s_1$ on the ghosts $\eta^{\alpha}.$
Since the vector fields $\{ X_a \}$ satisfy the Jacobi identity it follows from a computation similar to the one of Lemma (2.2) that 
$d^2 \eta^a =0.$
 Combined with the facts that $d^2=-[\delta, s_1]$ and that  $s_1\eta^{\alpha}$ has anti-ghost number one we have 
\begin{eqnarray}
0= -(\delta s_1 + s_1 \delta)\eta^\alpha =-\delta s_1 \eta^\alpha \\
=-\delta(C_{abc}^{d\alpha}\eta^a \eta^b \eta^c P_d)=(C_{abc}^{d\alpha}\eta^a \eta^b \eta^c G_d).
\end{eqnarray}
Therefore  $C_{abc}^{d\alpha}\eta^a \eta^b \eta^c G_d=0$ and consequently  we {\it can choose}  $C_{abc}^{d\alpha}=0.$  It follows that 
 $s_1 \eta^\alpha=0. $
 
At this point we perform some calculations  which are necessary to compute $ s_1 P_a.$
Since $d$ increases the ghost number by one we can write $dP_a=\eta^c C_{ca}^b P_b$ and 
\begin{eqnarray}
d^2 P_a=d(dP_a)=d(\eta^c C_{ca}^b P_b) 
=(d\eta^c) C_{ca}^b P_b-\eta^c d(C_{ca}^b P_b) \nonumber \\
=(-\frac{1}{2}C_{de}^c \eta^d \eta^e) P_b C_{ca}^b -\eta^c(\rho_d (C_{ca}^b )\eta^d P_b + C_{ca}^b dP_b) \nonumber \\
=-\frac{1}{2}C_{de}^c C_{ca}^b\eta^d \eta^e P_b -\rho_d (C_{ca}^b) \eta^c \eta^d P_b -C_{ca}^b \eta^c (\eta^e C_{eb}^d P_d) \nonumber \\
=(-\frac{1}{2}C_{ce}^b C_{ba}^d -C_{ca}^b C_{eb}^d -\rho_e(C_{ca}^d))\eta^c \eta^e P_d \label{l}
\end{eqnarray}
 
Since $s_1$ increases the anti-ghost number by one we can write $s_1 P_a=C_{cda}^{ef}\eta^c \eta^d P_e P_f .$
It follows that 
\begin{eqnarray}
\delta s_1 P_a=\delta(C_{cda}^{ef}\eta^c \eta^d P_e P_f )\\ =C_{cda}^{ef}\eta^c \eta^d((\delta P_e)P_f-P_e \delta P_f)\\
=C_{cda}^{ef}\eta^c \eta^d(-G_e P_f+G_f P_e)\\=-C_{cda}^{ef} G_e \eta^c \eta^d P_f +C_{cda}^{ef} G_f \eta^c \eta^d P_e \\
=2C_{cda}^{ef}\eta^c \eta^d P_e G_f
\end{eqnarray}
and 
\begin{eqnarray}
s_1 \delta P_a=s_1 (-G_a)=-s_1 G_a\\ =-(\rho_{cd}^e G_a))\eta^c \eta^d P_e
\end{eqnarray}
 Thus 
\begin{eqnarray}
(\delta s^1 + s^1 \delta)P_a=(2C_{cda}^{ef}G_f-\rho_{cd}^e (G_a))\eta^c \eta^d P_e\\
=(2C_{cea}^{df}G_f-\rho_{ce}^d (G_a))\eta^c \eta^e P_d \label{m}
\end{eqnarray}
Since $d^2=-[\delta, s^1]$,by comparing (\ref{l}) and ( \ref{m})
we have
\begin{eqnarray}
C_{cea}^{df}=\frac{1}{2G_f}(\rho_{ce}^d (G_a)+\rho_e(C_{ca}^d)+\frac{1}{2}C_{ce}^b C_{ba}^d +C_{ca}^b C_{eb}^d )
\end{eqnarray}
whenever $G_f $ is not zero.

We conclude that
\begin{eqnarray}
s_1 P_a=\frac{1}{2G_f}(\rho_{ce}^d (G_a)+\rho_e(C_{ca}^d)+\frac{1}{2}C_{ce}^b C_{ba}^d +C_{ca}^b C_{eb}^d )\eta^c \eta^e P_d P_f
\end{eqnarray}

{\bf In the last few paragraphs we have uniquely determined the action of $s_1$
on the generators of the BRST complex. We have found that $s_1$ is zero on all the ghost variables  and we have determined the value of $s_1$ on all the  anti-ghosts by the last rather complicated formula. These calculations determine the value of $s_1$ on the entire complex by the Leibniz formula.}

\section{The Maurer-Cartan structure of BRST differential under the reducible constraints}
In the last section we dealt only with irreducible constraints.
In this section, we will generalize some of our results  to  include  
systems  of reducible constraints. To achieve that,we   introduce the concept of an $n$-reducible complex as follows.
\begin{definition}
 Let $\Omega^*=\oplus_{n=0}^{\infty} \Omega^n $ be a graded algebra and assume that $\Omega^0$ is an algebra
 such that 
$\Omega^*$ is a $\Omega^0$-module.Let $A^p=\oplus_{n=0}^p \Omega^n .$ If $A^p$ is a finitely generated $\Omega^0$-module
and each component $\Omega^k (k>p) $ is  generated by $A^p ,$ We call the complex $\Omega^*$ a $p$th-reducible complex.
\end{definition}
We are  interested in investigating  differentials on $p$th-reducible complexes. First consider some examples of $p$th-reducible complexes.

\bigskip
\noindent{\bf Example 1.} Let $R^n$ be $n$-dimensional Euclidean space and $\Omega^k(R^n)$ be the space of $k$-forms. Since $R^n$ has a global coordinate
chart $(x^1,\cdots,x^n)$, every $k$-form can be written as 
\begin{eqnarray}
\omega= \omega_{i_1\cdots i_k} dx^{i_1}\wedge \cdots \wedge dx^{i_k}
\end{eqnarray}
where $\omega_{i_1\cdots i_k}$ are smooth functions which belong to the space of $0$-forms $\Omega^0 (R^n) $,
and where the $1$-forms $dx^1,\cdots , dx^n $  generate all differential forms in $\Omega^k (R^n) $ over $\Omega^0(R^n)$ for $ k\geq 1.$
The complex $\Omega^* (R^n)=\oplus_{k=0}^n \Omega^k (R^n) $ is then a $1$-reducible complex. An 
exterior differential $d$ can be defined as follows:
\begin{eqnarray}
df=(\partial_i  f) dx^i \\
d(dx^i)=0
\end{eqnarray}
where $ f\in \Omega^0 (R^n) ,dx^i \in \Omega^1(R^n)$. One then extends the definition  above to the entire space
 $\Omega^*(R^n)$ via  the Leibniz formula.

\bigskip
\noindent{\bf Example 2 (Chevalley-Eilenberg cohomology).}
Let a pair $(d,A)$ be Chevalley-Eilenberg differential and its  related complex 
$A={\cal A}\otimes C[\eta^1,\eta^2,\cdots ]$   discussed earlier.  Obviously, the graded algebra $A$ is generated by
ghosts $\eta^1,\eta^2,\cdots $ over  underlying algebra ${\cal A}$ and consequently  the complex $A$ is a 1th-reducible complex.

\bigskip
\noindent {\bf Example 3 (Auxillary Differential $\Delta $).}
Typical $k$-th reducible complexes arise from BRST operators in case the system is subject to  $(k-1)$th-reducible constraint conditions.
 
In BRST theory, a specific differential  called the auxillary differential $\Delta$ is introduced
to deal with  longitudinal differentials defined on constraint surfaces subject to higher order reducibility  conditions.
Let $\Sigma$ be a constraint surface and $\{ \eta^{a_0} \}$ be the original ghosts (see \cite{HT} page 217-218).  

Assume that 
$\Sigma$ is defined by constraints which satisfy  $k$-th order reducibility  
conditions. In this case the reducibility  of the constraints can be writen 
as
\begin{eqnarray}
Z_{a_k}^{a_{k-1}} 
Z_{a_{k-1}}^{a_{k-2}}=(-1)^{\epsilon_{a_{k-2}}C_{a_k}^{a_{k-2},a_0}}
G_{a_0}\\
k=1,\cdots, L, \quad \quad a_k =1,\cdots,m_k
\end{eqnarray}
for some $m_k$ and appropriate functions $ Z_{a_k}^{a_{k-1}}.$(see \cite{HT} page 210)

Introduce higher order ghost variables $\eta^a$ along with a differential $\Delta $ as follows:
\begin{eqnarray}
  puregh (\eta^{a_k})=k+1 , \quad \quad  \epsilon (\eta^{a_k})=\epsilon_{a_k}+k+1 \\
\Delta F=0, \quad  \quad \Delta \eta^{a_k}=\eta^{a_{k+1}} Z_{a_{k+1}}^{a_k}
  (-1)^{\epsilon_{a_k}+k+1 }
\end{eqnarray}
where $F$ is an arbitary function on the constraint surface. To complete the definition of the
auxillary differential $\Delta,$  one needs to introduce an auxillary grading as follows:
\begin{eqnarray}
 aux(z^A)=0=aux(\eta^{a^0}), \quad  aux(\eta^{a^k})=k
\end{eqnarray}
One then has that $aux(\Delta)=1$ and that $ puregh(A)=aux(A)+deg(A) $. The complex
$\Omega = C^{\infty}(\Sigma)\otimes C[\eta^{a_0},\eta^{a_1},\cdots, \eta^{a_k}] $ is then a $(k+1)th$-reducible complex.

At this point, we characterize the differential on an arbitrary $n$th-reducible complex and thereby generalize  
Theorem 1.  We adopt the convention of the left action for $d.$ Obviously, a differential $d$ 
on a reducible complex $\Omega^*$ is uniquely and totally determined  by its values on the
space $  A^p=\Omega^0 \oplus \bar A^p $ and the Leibniz rule, where $\bar A^p =\oplus_{n=1}^{ p} \Omega^n. $
In order to see this in more detail we first assume that the finitely generated $\Omega^0$ module $\bar A^p$
has a finite basis denoted by $( \omega _n^i)_{1\leq n \leq p} $ where the sub-index means $\omega_n^i \in \Omega^n$
Notice that the basis $( \omega _n^i)$  generates all the elements of every component $\Omega^k$ for $k>p$.
With this notation, we define a  differential $d$ on the space $A^p$
as follows :
\begin{eqnarray}
df=(\rho_j^1 f)\omega_1^j  \label{o1}\\ 
d\omega_m^i=-\frac{1}{2} \Sigma_{1\leq n \leq m}   C_{jkn}^i \omega_{m-n+1}^j \omega_n^k  \label {o}
\end{eqnarray}
where $f \in \Omega^0 $ and the $\rho_j^1$ are derivations of $\Omega^0$. If the complex $\Omega^*$ is a
$1th$-reducible complex, the basis has only $\omega_1^i$ type. then the formulas  (\ref{o1}) and (\ref{o})
reduce to 
\begin{eqnarray}
df=(\rho_j f)\omega_1^j \\ d\omega_1^i =-\frac{1}{2} C_{jk}^i \omega_1^j \omega_1^k   
\end{eqnarray}
Leibniz
which is analogous to the definition of the Chevalley-Eilenberg differential.

The definition of the Koszul-Tate differential $\delta$ is then modified to reflect the 
reducibility conditions above.
In the reducible case the longitudinal differential is not nilpotent on the space 
of
$C^{\infty}(\Sigma)\otimes C[\eta^a],$ but is nilpotent on the subalgebra of 
longitudinal forms. In order to overcome this defect, an equivalent differential $D$ is 
introduced
such that $H^*(D)=H^*(d)$. At this point one has  ``differentials" $\delta$ and $D$ on the 
extended space
$\Omega^*=C[P_{a_0},P_{a_1}.\cdots]\otimes C^{\infty}(\Sigma) \otimes 
C[\eta^{a_1},\eta^{a_2},\cdots]$
and it is possible to show that  exists a BRST differential $S=\delta +D+s_1 +\cdots$ defined on 
the complex $\Omega^*.$
It can be seen that  the longitudinal  complex 
$C^{\infty}(\Sigma)\otimes C[\eta^a]$ and the BRST complex $\Omega^*$ are 
both $n$th-reducible complexes for appropriate $n.$ Therefore
we obtain the following generalization of Theorem 1.
\begin{theorem}
If the constraints are Bosonic and reducible the BRST differential has a   
structure similar to that of the longitudinal differential.
\end{theorem}

\newpage

\begin{thebibliography}{10}

\bibitem{Samer} S. AL-Ashhab,  {\em A class of strongly homotopy Lie 
algebras with
simplied sh-Lie structures}, Los Alamos Archive,  math.RA/0308160

\bibitem{BBvD} F.A.Berends,G.J.H.Burgers and H.van Dam, Nucl.phys. B 
{\bf 260}
(1985), 295

\bibitem{B} G.Barnich, {\em Brackets in the jet-bundle approach to field 
theory},
Proceedings of the conference on Secondary Calculus and
Cohomological Physics, Contemp. Math. 219 (1998) 17-27.

\bibitem{BH} G.Barnich and M.Henneaux, Phys,Rev,Lett, {\bf 72} (1993), 1588

\bibitem{BFLS} G. Barnich,~R. Fulp,~T. Lada,~J. Stasheff,  {\em The sh 
Lie structure
of Poisson brackets in field theory}, Commun. Math. Phys. {\bf 
191}(1998),585-601

\bibitem{Brandt}F.Brandt, {\em Local BRST cohomology in the antifield 
formalism},
unpublished lecture notes

\bibitem {G} M.Gerstenhaber,  {\em The cohomology structure of an 
associate ring}
Annals of Mathematics {\bf 78} 267-288

\bibitem{HG} M.Hazewinkel and M.Gerstenharber, {\em Deformation theory 
of Lie
algebras and structures and applications},
NATO Series C,Mathematics and Physics  Science {\bf 24}

\bibitem{HT} M.~Henneaux and C.~Teitelboim, {\em Quantization of {G}auge 
{S}ystems},
 Princeton Univ. Press, 1992.

\bibitem{KV} I.S. Krasilshchik and A.M. Vinogradov,  {\em Symmetries and
conservation laws for differential equations of mathematical physics},  
Translations
of mathematical monographs {\bf 182,} AMS (1999)

\bibitem{ls}T.~Lada and J.D. Stasheff, {\em Introduction to sh {L}ie 
algebras for
physicists}, Intern'l J. Theor. Phys. {\bf 32} (1993), 1087--1103.

\bibitem{Olver}P. Olver {\em Applications of Lie groups to differential 
equations}
Graduate texts in Mathematics, Vol 107, Berlin-Heidelberg-New York: 
Springer-Verlag,
1986

\bibitem{SS} M.Schlessinger,J.Stasheff,  {\em The {L}ie algebra 
structure and
tangent cohomology and deformation}, J.Pure Appl. Algebra {\bf 89} 
(1993), 231-235


\bibitem{s} J.Stasheff, {\em Deformation Theory and the 
Batalin-Vilkovisky Master
Equation},
Proceedings of the Conference on Deformation Theory, etc. Ascona, 
Switzerland (1996)
q-alg/9702012
  
 



\end{thebibliography}

\ifx\undefined\bysame
\newcommand{\bysame}{\leavevmode\hbox to3em{\hrulefill}\,}
\fi

\end{document}